\documentclass[letter,twocolumn]{jpsj3}
\usepackage{txfonts}

\title{
  Anisotropic Superconductivity Emerging from the Orbital Degrees of Freedom\\
  in a $\Gamma_3$ Non-Kramers Doublet System
}

\author{Katsunori Kubo}
\inst{
Advanced Science Research Center, Japan Atomic Energy Agency,
  Tokai, Ibaraki 319-1195, Japan
} 

\abst{
We study superconductivity in a three-orbital model for $f^2$ ions
with the $\Gamma_3$ crystalline electric field (CEF) ground state.
An antiferromagnetic interaction
between the $\Gamma_7$ and $\Gamma_8$ orbitals
is introduced to stabilize the $\Gamma_3$ CEF state.
This interaction also works as an on-site attractive interaction
for spin-singlet pairing between electrons in these orbitals.
The interorbital pairing state
composed of the $\Gamma_7$ and $\Gamma_8$ orbitals
on the same site has the $E_g$ symmetry.
Indeed, by applying the random phase approximation,
we find that the $E_g$ spin-singlet superconducting state is realized
over a wide parameter range.
}

\begin{document}
\maketitle

Unconventional superconductivity~\cite{Sigrist1991,Norman2011},
such as
in cuprate high-temperature superconductors~\cite{Pickett1989,Dagotto1994},
Fe-based superconductors~\cite{Ishida2009,Stewart2011},
and heavy-fermion materials~\cite{Onuki2004,Lohneysen2007},
is often observed near the antiferromagnetically ordered phase.
The relation between superconductivity and magnetism
has thus been a central issue in condensed matter physics.
In particular, the magnetic-fluctuation-mediated
superconducting mechanism has been widely discussed.

In addition,
the orbital degrees of freedom may play an important role
in superconductivity in orbitally degenerate systems.
Indeed, there have been suggestions that orbital fluctuations
are important in Fe-based superconductors.
However, the presence of both orbital and spin degrees of freedom
has made it difficult to discern
the specific role of the orbital degrees of freedom.

On the other hand, in $f$-electron systems,
an ion with an even number of $f$ electrons
has the $\Gamma_3$ non-Kramers doublet level
under a cubic crystalline electric field (CEF).
The $\Gamma_3$ state has the same symmetry as the spinless $e_g$ electron,
and possesses quadrupole and octupole moments
but no dipole moment.
Thus, the $\Gamma_3$ system can be regarded as an ideal system
for investigating orbital physics,
and may provide a route to unconventional superconductivity
other than the spin-fluctuation mechanism.

Superconductivity was observed in
PrT$_2$X$_{20}$ (where T denotes a transition metal and X is Zn or Al)
in which the CEF ground state of the $f^2$ electronic configuration
in a Pr$^{3+}$ ion
is the $\Gamma_3$ doublet~\cite{Onimaru2011,Sakai2011,Ishii2011,Sato2012,
  Ishii2013,Iwasa2013,Hamamoto2017}.
(Strictly, in PrRh$_2$Zn$_{20}$,
the CEF ground state is the $\Gamma_{23}$ doublet
owing to symmetry lowering at the Pr site that is
induced by a structural transition~\cite{Iwasa2013}.)
In PrIr$_2$Zn$_{20}$~\cite{Onimaru2010,Onimaru2011,Ishii2011,Iwasa2017}
and PrV$_2$Al$_{20}$~\cite{Sakai2011,Tsujimoto2014},
superconductivity is observed
below the antiferroquadrupole ordering temperature.
The order parameter for antiferroquadrupole ordering
in PrIr$_2$Zn$_{20}$ was determined to be $O^2_2=x^2-y^2$~\cite{Iwasa2017}.
In PrTi$_2$Al$_{20}$,
superconductivity appears below the ferroquadrupole ordering
temperature of $O^0_2=3z^2-r^2$~\cite{Sakai2011,Sato2012,Ito2011,Sakai2012,
  Matsubayashi2012,Matsubayashi2014,Taniguchi2016}.
In PrRh$_2$Zn$_{20}$, superconductivity occurs simultaneously
with antiferroquadrupole ordering~\cite{Onimaru2012,Ishii2013}.
In any case, superconductivity is realized
inside the quadrupole ordered phase.
The relation between superconductivity
and the quadrupole degrees of freedom has therefore attracted much attention.

To discuss superconductivity with quadrupole or orbital degrees of freedom,
it may be useful to consult two-orbital models.
In two-orbital models,
there is an interesting possibility to realize anisotropic superconductivity
originating from the orbital anisotropy~\cite{Kubo2007,Hattori2017}.
For example, we obtained $d$-wave spin-triplet superconductivity
in a model for $e_g$ orbitals on a square lattice~\cite{Kubo2007}.
However, it was shown that
a two-orbital model is inadequate for describing
the multipole degrees of freedom
in the $\Gamma_3$ CEF state~\cite{Kubo2017,Kubo2018}.
For example, the intermediate $f^3$ state is always the $\Gamma_8$ state
in the two-orbital model for the $\Gamma_3$ doublet,
but for a realistic parameter set to realize the $f^2$-$\Gamma_3$ state
in a local model considering all the $f$-electron orbitals,
the $f^3$ ground state is the $\Gamma_6$ state~\cite{Hotta2006}.
It is therefore necessary to look beyond the two-orbital model.
Indeed, we have found that a three-orbital model can remedy the above shortcomings~\cite{Kubo2017}.

\begin{figure}
  \includegraphics[width=1.0\linewidth]
  {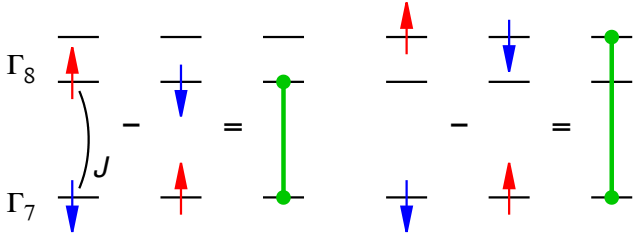}
  \caption{\label{level_scheme}
    (Color online)
    Electron configurations of the $\Gamma_3$ doublet.
    The bold lines denote spin singlets
    composed of the $\Gamma_7$ and $\Gamma_8$ orbitals.
  }
\end{figure}

The present study considers the three-orbital model
for the $\Gamma_3$ CEF state~\cite{Kubo2017}
by applying the random-phase approximation (RPA)~\cite{Takimoto2000},
and clarifies the characteristics of superconductivity
in the $\Gamma_3$ systems.
In the present model,
the $\Gamma_3$ doublet is composed of the two singlets between
the $\Gamma_7$ and $\Gamma_8$ orbitals (Fig.~\ref{level_scheme}).
The interaction that stabilizes the $\Gamma_3$ doublet
also works as an on-site attractive interaction
for the spin-singlet pairing between the electrons in these orbitals.
The orbital symmetry can be rewritten as
$\Gamma_7=\Gamma_2 \times \Gamma_6$ and
$\Gamma_8=\Gamma_3 \times \Gamma_6$,
where $\Gamma_6$ describes the Kramers -- or spin -- degeneracy.
Thus, the interorbital spin-singlet pairing state, composed
of the $\Gamma_7$ and $\Gamma_8$ orbitals on the same site has
the $E_g$ $(=\Gamma_3=\Gamma_2 \times \Gamma_3)$ symmetry.
It is therefore natural to expect $d$-wave superconductivity in this model.

This study
considers the $f$-electron states
with total angular momentum $j=5/2$
as one-electron states.
These states split into the $\Gamma_7$ and $\Gamma_8$ states
in a cubic CEF.
The $\Gamma_7$ states at site $\mib{r}$ are given by
$c^{\dagger}_{\mib{r} 7 \uparrow} |0 \rangle
=
(1/\sqrt{6})
(a^{\dagger}_{\mib{r} 5/2}
-\sqrt{5} a^{\dagger}_{\mib{r} -3/2})|0 \rangle$
and
$c^{\dagger}_{\mib{r} 7 \downarrow} |0 \rangle
=
(1/\sqrt{6})
(a^{\dagger}_{\mib{r} -5/2}
-\sqrt{5} a^{\dagger}_{\mib{r} 3/2})|0 \rangle$,
where
$a^{\dagger}_{\mib{r} j_z}$ is the creation operator
for the electron with $j_z$ as the $z$-component of the total momentum at $\mib{r}$,
and $|0\rangle$ denotes the vacuum state.
The $\Gamma_8$ states are given by
$c^{\dagger}_{\mib{r} \alpha \uparrow} |0 \rangle
= 
(1/\sqrt{6})
( \sqrt{5}a^{\dagger}_{\mib{r} 5/2}
+a^{\dagger}_{\mib{r} -3/2})|0 \rangle$,
$c^{\dagger}_{\mib{r} \alpha \downarrow} |0 \rangle
= 
(1/\sqrt{6})
( \sqrt{5}a^{\dagger}_{\mib{r} -5/2}
+a^{\dagger}_{\mib{r} 3/2} )|0 \rangle$,
$c^{\dagger}_{\mib{r} \beta \uparrow} |0 \rangle
=
a^{\dagger}_{\mib{r} 1/2}|0 \rangle$,
and
$c^{\dagger}_{\mib{r} \beta \downarrow} |0 \rangle
=
a^{\dagger}_{\mib{r} -1/2}|0 \rangle$.
In these states,
$\sigma = \uparrow$ or $\downarrow$ denotes the Kramers degeneracy
of the one-electron states.
Although this is not a real spin, owing to spin-orbit coupling,
we may nonetheless henceforth call it spin for simplicity.

To realize the $\Gamma_3$ state as the ground state of an $f^2$ ion,
we consider an antiferromagnetic interaction $J$ between electrons
in the $\Gamma_7$ and $\Gamma_8$ orbitals~(Fig.~\ref{level_scheme}).
While it may be possible to derive $J$ by taking account of the effects
of the higher-energy $j=7/2$ states~\cite{Hotta2006},
we introduce it here phenomenologically
to realize the $\Gamma_3$ state.
This interaction favors spin singlets
between these orbitals.
These singlets are the main components of the $\Gamma_3$ doublet
in realistic situations~\cite{Hotta2006,Kubo2017}.
When $J$ is large enough,
the ground state among the $f^3$ states
is the $\Gamma_6$ doublet~\cite{Kubo2017}.
The drawback of the two-orbital model
discussed above is thus eliminated.

The model Hamiltonian is given by
\begin{equation}
  H
  =\sum_{\mib{r},\mib{\mu},\gamma,\gamma^{\prime}}
  c^{\dagger}_{\mib{r}+\mib{\mu} \gamma}
  t^{\mib{\mu}}_{\gamma \gamma^{\prime}}
  c_{\mib{r} \gamma^{\prime}}
  +J \sum_{\mib{r}} \mib{s}_{\mib{r}7} \cdot \mib{s}_{\mib{r}8},
\end{equation}
where
$t^{\mib{\mu}}_{\gamma \gamma'}$ is the hopping integral, with
the vector $\mib{\mu}$ connecting nearest-neighbor sites
and
$\gamma=(\tau,\sigma)$ with $\tau=\alpha$, $\beta$, or 7.
$\mib{s}_{\mib{r} 7}=(1/2) \sum_{\sigma \sigma'}
c^{\dagger}_{\mib{r} 7 \sigma} \mib{\sigma}_{\sigma \sigma'} c_{\mib{r} 7 \sigma'}$
and
$\mib{s}_{\mib{r} 8}=(1/2) \sum_{\nu \sigma \sigma'}
c^{\dagger}_{\mib{r} \nu \sigma} \mib{\sigma}_{\sigma \sigma'} c_{\mib{r} \nu \sigma'}$
with $\nu=\alpha$ or $\beta$.
$\mib{\sigma}$ are the Pauli matrices.
Note that
$t^{\mib{\mu} *}_{\gamma \gamma'}=t^{-\mib{\mu}}_{\gamma' \gamma}$,
since $H$ is Hermitian.

With regard to the kinetic energy term,
as one of the simplest possible models,
we consider only $f$-electron hopping
through $\sigma$ bonding $(ff\sigma)$
on a simple cubic lattice.
In this case, the hopping integrals are nonzero
only between the $\Gamma_8$ orbitals
and they can be expressed as $4\times4$ matrices~\cite{Hotta2003,Kubo2005PRB}.
The hopping integrals are given by
$t^{(1,0,0)} = (\tilde{1}-\tilde{\eta}^+)t$,
$t^{(0,1,0)} = (\tilde{1}-\tilde{\eta}^-)t$,
and
$t^{(0,0,1)} = (\tilde{1}-\tilde{\tau}^z)t$,
where
$\tilde{1}_{\nu \sigma; \nu^{\prime} \sigma^{\prime}}
= \delta_{\nu \nu^{\prime}} \delta_{\sigma \sigma^{\prime}}$,
$\tilde{\mib{\tau}}_{\nu \sigma; \nu^{\prime} \sigma^{\prime}}
= \mib{\sigma}_{\nu \nu^{\prime}} \delta_{\sigma \sigma^{\prime}}$,
and
$\tilde{\eta}^{\pm}
= (\pm \sqrt{3}\tilde{\tau}^x-\tilde{\tau}^z)/2$.
We have set the lattice constant to unity
and $t=3(ff\sigma)/14$.
The bandwidth is $W=12t$.

When strong electron correlations are properly included
for the $f^2$ case (that is, with two electrons per site in this model),
the electron number in each of the $\Gamma_7$ and $\Gamma_8$ levels
should be nearly equal to one:
$n_7=
\langle \sum_{\sigma}c^{\dagger}_{\mib{r} 7 \sigma}c_{\mib{r} 7 \sigma} \rangle \simeq 1$
and
$n_8=\langle \sum_{\nu \sigma}
c^{\dagger}_{\mib{r} \nu \sigma}c_{\mib{r} \nu \sigma} \rangle \simeq 1$,
where $\langle \cdots \rangle$ denotes the expectation value.
Such strong correlation effects can be included partly within the RPA by fixing $n_7$ and $n_8$, via the independent tuning of the chemical potentials
for these orbitals.

The gap equation is expressed as~\cite{Takimoto2000,Kubo2006JPSJ}
\begin{equation}
  \begin{split}
    &\lambda \Delta^{\xi}_{\tau_1 \tau_2}(\mib{k})\\
    =&
    -\frac{1}{N}\sum_{\mib{k}' \tau_3 \tau_4 \tau'_1 \tau'_2}
    V^{\xi}_{\tau_1 \tau_3; \tau_4 \tau_2}(\mib{k}-\mib{k}')
    \phi_{\tau_3 \tau'_1; \tau_4 \tau'_2}(\mib{k}')
    \Delta^{\xi}_{\tau'_1 \tau'_2}(\mib{k}'),
  \end{split}
\end{equation}
where $N$ is the number of lattice sites
and
$\Delta^{\text{s}}(\mib{k})$ [$\Delta^{\text{t}}(\mib{k})$]
is the gap function for spin-singlet [-triplet] pairing.
The eigenvalue $\lambda$ reaches unity
at the superconducting transition temperature $T_{\text{c}}$.
The pair-correlation function $\phi(\mib{k})$ is given by
\begin{equation}
  \phi_{\tau_1 \tau'_1; \tau_2 \tau'_2}(\mib{k})
  =
  T\sum_{i\epsilon_n}
  G_{\tau_1 \tau'_1}(\mib{k},i\epsilon_n)G_{\tau_2 \tau'_2}(-\mib{k},-i\epsilon_n),
  \label{eq:phi}
\end{equation}
where
$T$ is the temperature and
$G_{\tau_1 \tau_2}(\mib{k},i\epsilon_n)$ is the non-interacting Green's function.
The pairing interactions are written as
\begin{align}
  V^{\text{s}}(\mib{q})
  &=
  \frac{3}{2}\left[
    U^{\text{s}}\chi^{\text{s}}(\mib{q})U^{\text{s}}+\frac{U^{\text{s}}}{2} \right]
  -\frac{1}{2}\left[
    U^{\text{c}}\chi^{\text{c}}(\mib{q})U^{\text{c}}-\frac{U^{\text{c}}}{2} \right],
  \label{eq:Vs}\\
  V^{\text{t}}(\mib{q})
  &=
  -\frac{1}{2}\left[
    U^{\text{s}}\chi^{\text{s}}(\mib{q})U^{\text{s}}+\frac{U^{\text{s}}}{2} \right]
  -\frac{1}{2}\left[
    U^{\text{c}}\chi^{\text{c}}(\mib{q})U^{\text{c}}-\frac{U^{\text{c}}}{2} \right],
\end{align}
where the spin and charge susceptibilities are
\begin{equation}
  \chi^{\text{s/c}}(\mib{q})
  = \chi(\mib{q})\left[1 \mp U^{\text{s/c}}\chi(\mib{q})\right]^{-1},
\end{equation}
with
\begin{equation}
  \chi_{\tau_1 \tau'_1; \tau_2 \tau'_2}(\mib{q})
  =
  -\frac{T}{N}\sum_{\mib{k}, i\epsilon_n}
  G_{\tau_1 \tau_2}(\mib{k}+\mib{q},i\epsilon_n)
  G_{\tau'_2 \tau'_1}(\mib{k},i\epsilon_n).
  \label{eq:chi}
\end{equation}
The matrices $U^{\text{s}}$ and $U^{\text{c}}$ are defined as
$U^{\text{s}}_{7 \nu ;7 \nu}=U^{\text{s}}_{\nu 7;\nu 7}=U'_{78}\equiv-J/4$,
$U^{\text{s}}_{7 7; \nu \nu}=U^{\text{s}}_{\nu \nu; 7 7}=J_{78}\equiv-J/2$,
$U^{\text{c}}_{7 \nu ;7 \nu}=U^{\text{c}}_{\nu 7;\nu 7}=-U'_{78}+2J_{78}$,
$U^{\text{c}}_{7 7; \nu \nu}=U^{\text{c}}_{\nu \nu; 7 7}=2U'_{78}-J_{78}$,
and zero otherwise.
In the evaluation of Eqs.~\eqref{eq:phi} and \eqref{eq:chi},
the summation over the fermion Matsubara frequency $\epsilon_n$
can be executed analytically.

The multipole operators can be expressed in the form
\begin{equation}
  O_{\mib{r}}
  =
  \sum_{\gamma \gamma'}
  c^{\dagger}_{\mib{r} \gamma} \tilde{O}_{\gamma \gamma'}c_{\mib{r} \gamma'}.
\end{equation}
We normalize $\tilde{O}$
so that $\text{Tr} \tilde{O}^2=1$,
where Tr denotes the trace of the matrix.
Using linear-response theory,
we can calculate the multipole susceptibility~\cite{Kubo2006JPSJS}.
The susceptibility for a magnetic [electric] multipole moment
can be expressed
by using $\chi^{\text{s}}(\mib{q})$ [$\chi^{\text{c}}(\mib{q})$].
If we determine the multipole ordering temperature
from the divergence of the corresponding susceptibility $\chi$,
we will always find superconducting instability
before multipole ordering takes place,
owing to the enhancement of the pairing interaction.
The root of this unrealistic consequence
is the ignorance of the self-energy within RPA~\cite{Shimahara1988}.
In this study, we identify the multipole ordering temperature
as the temperature where $\chi$ reaches a particular threshold.

In the following,
we show the results for a simple cubic lattice
of size $N=32 \times 32 \times 32$.
We also performed calculations for a $16 \times 16 \times 16$ lattice
and found the size dependence to be negligible
except for the dilute cases $n_8 \simeq 0$.
Concerning multipole ordering,
we define the transition temperature
by the condition $\chi=10/t$.
Then, we determine the highest transition temperature
among the multipole-ordering and superconducting transitions.
Having considered all the superconducting states,
we found that only the $E_g$ spin-singlet pairing state displays
the highest transition temperature.
For multipole ordering,
we consider the charge, dipole, quadrupole, and octupole moments.
Among them,
only the $\Gamma_{4}$ octupole ($\mib{T}^{\alpha}$) state is realized.
It has the same symmetry as the dipole moment,
and should appear simultaneously with the dipole moment in the ordered phase.
We found that the susceptibilities for $\mib{T}^{\alpha}$
and for the dipole moment diverge at the same temperature,
if we ignore the superconducting instability.
Thus, we determined the spin-density-wave (SDW) transition temperature
$T_{\text{SDW}}$ by using the susceptibility for $\mib{T}^{\alpha}$.

\begin{figure}
  \includegraphics[width=1.0\linewidth]
  {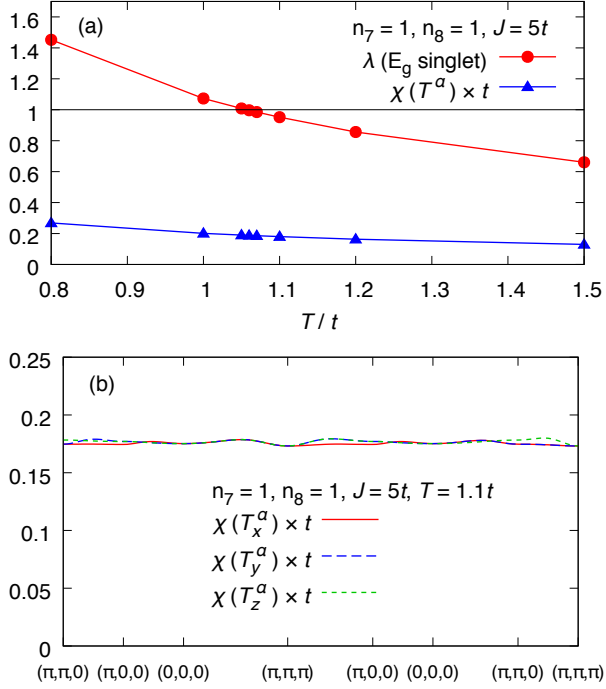}
  \caption{\label{lambda_chi}
    (Color online)
    (a) Temperature dependences
    of the eigenvalue $\lambda$ for the $E_g$ spin-singlet pairing
    and
    of the susceptibility $\chi$ for $\mib{T}^{\alpha}$ at $\mib{q}_{\text{max}}$
    for $n_7=n_8=1$ and $J=5t$.
    (b) Wave vector dependences of the susceptibilities
    for $n_7=n_8=1$, $J=5t$, and $T=1.1t$.
  }
\end{figure}

Figure~\ref{lambda_chi}(a) shows the temperature dependences
of the eigenvalue $\lambda$ for the $E_g$ spin-singlet superconductivity
and
of the susceptibility $\chi$ for $\mib{T}^{\alpha}$
at the wave vector $\mib{q}_{\text{max}}$,
where the susceptibility has the maximum value
for $n_7=n_8=1$ and $J=5t$.
The value of $\lambda$ reaches unity at $T_{\text{c}} \simeq 1.06 t$.
The susceptibility $\chi$ is not enhanced for this parameter set.
Thus, the superconductivity is not mediated by such multipole fluctuations
with a particular wave vector.
The fluctuations just above the transition temperature
depend weakly on the wave vector [Fig.\ref{lambda_chi}(b)],
that is, the local fluctuations may still be important for the emergence
of superconductivity.

\begin{figure}
  \includegraphics[width=1.0\linewidth]
  {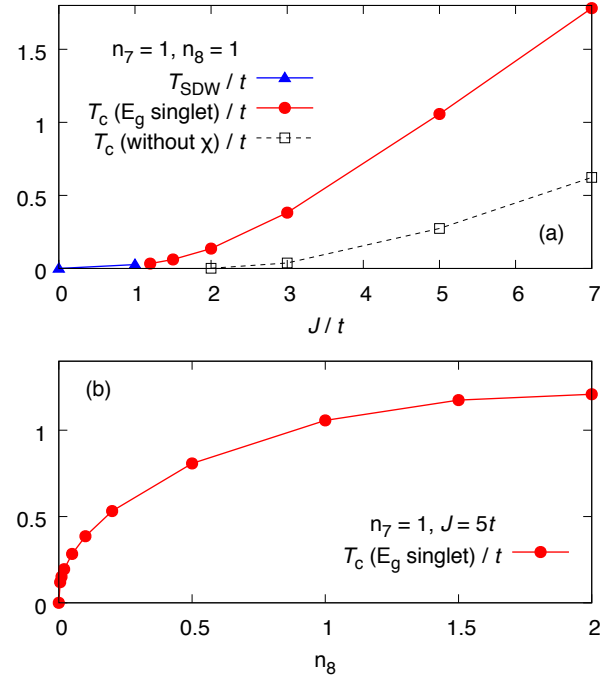}
  \caption{\label{Tc}
    (Color online)
    (a) $T_{\text{c}}$ for the $E_g$ spin-singlet pairing and $T_{\text{SDW}}$
    as functions of $J$
    for $n_7=n_8=1$.
    The open squares represent $T_{\text{c}}$ evaluated
    without $\chi^{\text{s}}(\mib{q})$ and $\chi^{\text{c}}(\mib{q})$
    in Eq.~\eqref{eq:Vs}.
    (b) $T_{\text{c}}$ for the $E_g$ spin-singlet pairing as a function of $n_8$
    for $n_7=1$ and $J=5t$.
  }
\end{figure}

Figure~\ref{Tc}(a) shows
$T_{\text{c}}$ for the $E_g$ spin-singlet pairing and $T_{\text{SDW}}$
as functions of the strength of the antiferromagnetic interaction $J$.
In the cases of weak interactions ($J \lesssim t$),
the SDW state is realized but its transition temperature $T_{\text{SDW}}$
is very low.
For stronger interactions ($J \gtrsim t$),
the $E_g$ spin-singlet superconducting state is realized.
The transition temperature $T_{\text{c}}$ increases with $J$.
In a wide parameter region $J \lesssim W=12t$,
where we can apply weak-coupling theory,
the $E_g$ spin-singlet superconducting state is realized.
Note that, for a larger $J$,
local spin-singlet formation
(which is beyond the scope of the present weak-coupling treatment)
should become strong,
leading to the suppression of superconductivity.
In Fig.~\ref{Tc}(a), we also show $T_{\text{c}}$ evaluated
without $\chi^{\text{s}}(\mib{q})$ and $\chi^{\text{c}}(\mib{q})$
in Eq.~\eqref{eq:Vs}.
In this case, $T_{\text{c}}$ becomes much lower.
We therefore recognize that the local fluctuations
$\sim \sum_{\mib{q}} \chi(\mib{q})$ are still important
for the realization of superconductivity
and that we cannot ignore $\chi(\mib{q})$,
even though $\chi(\mib{q}_{\text{max}})$ is not large.

To examine the stability of the $E_g$ spin-singlet superconducting state,
we varied the electron number $n_8$ in the $\Gamma_8$ level from unity.
For example,
we show the $n_8$ dependence of $T_{\text{c}}$ for $J=5t$
in Fig.~\ref{Tc}(b).
We find that
$T_{\text{c}}$ is always higher than for the other superconducting states
and $T_{\text{SDW}}$ for this parameter.
Thus, the $E_g$ spin-singlet state is not restricted to $n_8 \simeq 1$.

\begin{figure}
  \includegraphics[width=1.0\linewidth]
  {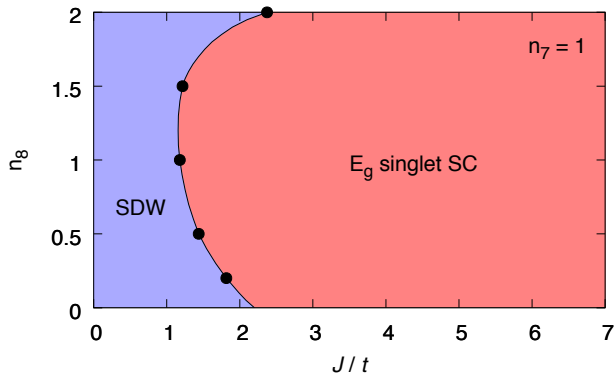}
  \caption{\label{phase_diagram}
    (Color online)
    Phase diagram for $n_7=1$.
  }
\end{figure}

By determining the highest transition temperature for each parameter set,
we constructed the phase diagram shown in Fig.~\ref{phase_diagram}.
The change in $J$ corresponds to the external hydrostatic pressure
and the change in $n_8$ is regarded as carrier doping.
In this phase diagram, the $E_g$ spin-singlet superconductivity 
is realized over a wide range of parameters.
This superconductivity would therefore be robust against
such perturbations.
Notably, this superconducting phase extends
far from the SDW phase
and does not require particular fluctuations.
Even in an unrealistic two-orbital model for the $\Gamma_3$ CEF state,
$E_g$ superconductivity was obtained,~\cite{Hattori2017}
indicating that $E_g$ superconductivity is relatively stable
in $\Gamma_3$ systems.

In summary,
we have investigated superconductivity
in the $\Gamma_3$ non-Kramers doublet system.
The $E_g$ spin-singlet pairing state
originating from the orbital degrees of freedom
is realized over a wide parameter range.
This superconductivity is not mediated by fluctuations
with a particular wave vector
and would therefore be stable against perturbations.

\begin{acknowledgment}
This work was supported by JSPS KAKENHI Grant Number
JP15K05191. 
\end{acknowledgment}


\end{document}